\documentclass[english,aps,preprint]{revtex4-1}
\pdfoutput=1
\usepackage[T1]{fontenc}
\usepackage[latin9]{inputenc}
\usepackage{amsmath}
\usepackage{graphicx}
\usepackage{amssymb}
\usepackage{esint}

\makeatletter
\@ifundefined{textcolor}{}
{%
 \definecolor{BLACK}{gray}{0}
 \definecolor{WHITE}{gray}{1}
 \definecolor{RED}{rgb}{1,0,0}
 \definecolor{GREEN}{rgb}{0,1,0}
 \definecolor{BLUE}{rgb}{0,0,1}
 \definecolor{CYAN}{cmyk}{1,0,0,0}
 \definecolor{MAGENTA}{cmyk}{0,1,0,0}
 \definecolor{YELLOW}{cmyk}{0,0,1,0}
 }

\makeatother
\usepackage{babel}
\begin{document}

\title{Effective Langevin equations for constrained stochastic processes}

\author{Satya N. Majumdar}
\affiliation{Laboratoire de Physique Th\'eorique et Mod\`eles Statistiques (UMR 8626 du CNRS),\\ 
Universit\'e Paris-Sud, B\^atiment 100, 91405 Orsay Cedex, France} 
\author{Henri Orland}
\affiliation{Institut de Physique Th\'eorique, CEA, IPhT \\
 CNRS, URA2306, \\
 F-91191 Gif-sur-Yvette, France\\
 and\\
Beijing Computational Science Research Center \\
No.3 HeQing Road, Haidian District\\
Beijing, 100084, China 
 }

\email{satya.majumdar@u-psud.fr, henri.orland@cea.fr}

\begin{abstract}
We propose a novel stochastic method to exactly generate Brownian paths conditioned
to start at an initial point and end at a given final point during
a fixed time $t_{f}$. These paths are weighted with a probability
given by the overdamped Langevin dynamics. We show how these paths
can be exactly generated by a local stochastic differential equation. The method is illustrated on the generation of Brownian bridges, Brownian meanders, Brownian excursions and constrained Ornstein-Uehlenbeck processes. In addition, we show how to solve this equation in the case of a general force acting on the particle. As an example, we show how to generate constrained path joining the two minima of a double-well. Our method allows to generate statistically independent paths, and is computationally very efficient.
\end{abstract}
\maketitle

\section{Introduction}
Even after more than hundred years since its introduction, Brownian motion has remained
a fundamental cornerstone of classical physics~\cite{brownian}. Simple Brownian motion and 
its variants
have found numerous applications not just in natural sciences such as physics, chemistry, biology etc.,
but also in man made subjects such as finance, economics, queueing theory, search processes,
and computer science amongst others (for reviews on
the subject see~\cite{usage,Duplantier,Yor,CDT,majumdar_review}). Some of 
these applications led to the studies of
the statistical properties of simple variants of Brownian motion, generally referred to
as `constrained Brownian motion', i.e, a free Brownian motion conditioned to satisfy
certain prescribed global constraints (see below for examples). More generally, one would also like
to study other stochastic processes (thus going beyond the simple Brownian motion) in presence of one or more
such global constraints. An interesting and challenging problem is to find
simple and efficient algorithms that generate these constrained paths, for Brownian motion
and other stochastic processes, with the correct statistical weight. 
Many of the concepts presented in this article through a physical presentation, can be found in the
mathematical literature, for example in ~\cite{rogers_williams}.

As a simple example, let us first start with a {\em free} Brownian motion $B(t)$ in one dimension, starting  
from the origin $B(0)=0$. The time evolution of $B(t)$ is governed by the simple Langevin equation
\begin{equation}
\frac{dB}{dt}= \eta(t)\,,
\label{brown_t}
\end{equation}
where $\eta(t)$ is a Gaussian white noise with zero mean $\langle \eta(t)\rangle =0$
and the correlator, $\langle \eta(t)\eta(t')\rangle=  2\,D\, \delta(t-t')$ ($D$ being the
diffusion constant). It is easy to generate a trajectory of this process numerically, simply
by simulating the time-discretized version of the Langevin equation
\begin{equation}
B(t+\Delta t) = B(t) + \sqrt{2\, D\, {\Delta t}}\, \xi_t\, ,
\label{brown_discrete}
\end{equation}
where the noise $\xi_t \equiv {\cal N}(0,1)$ is a standard normally
distributed random variable with zero mean and unit variance, 
drawn independently at each discrete time step. Three globally constrained variants (amongst others)
of this simple $1$-d Brownian motion that have been widely studied in the literature are
the so called (i) Brownian bridge (ii) Brownian excursion and (iii) Brownian meander that
we discuss next.

\vskip 0.3cm

\begin{figure}
\includegraphics[width=.49\hsize]{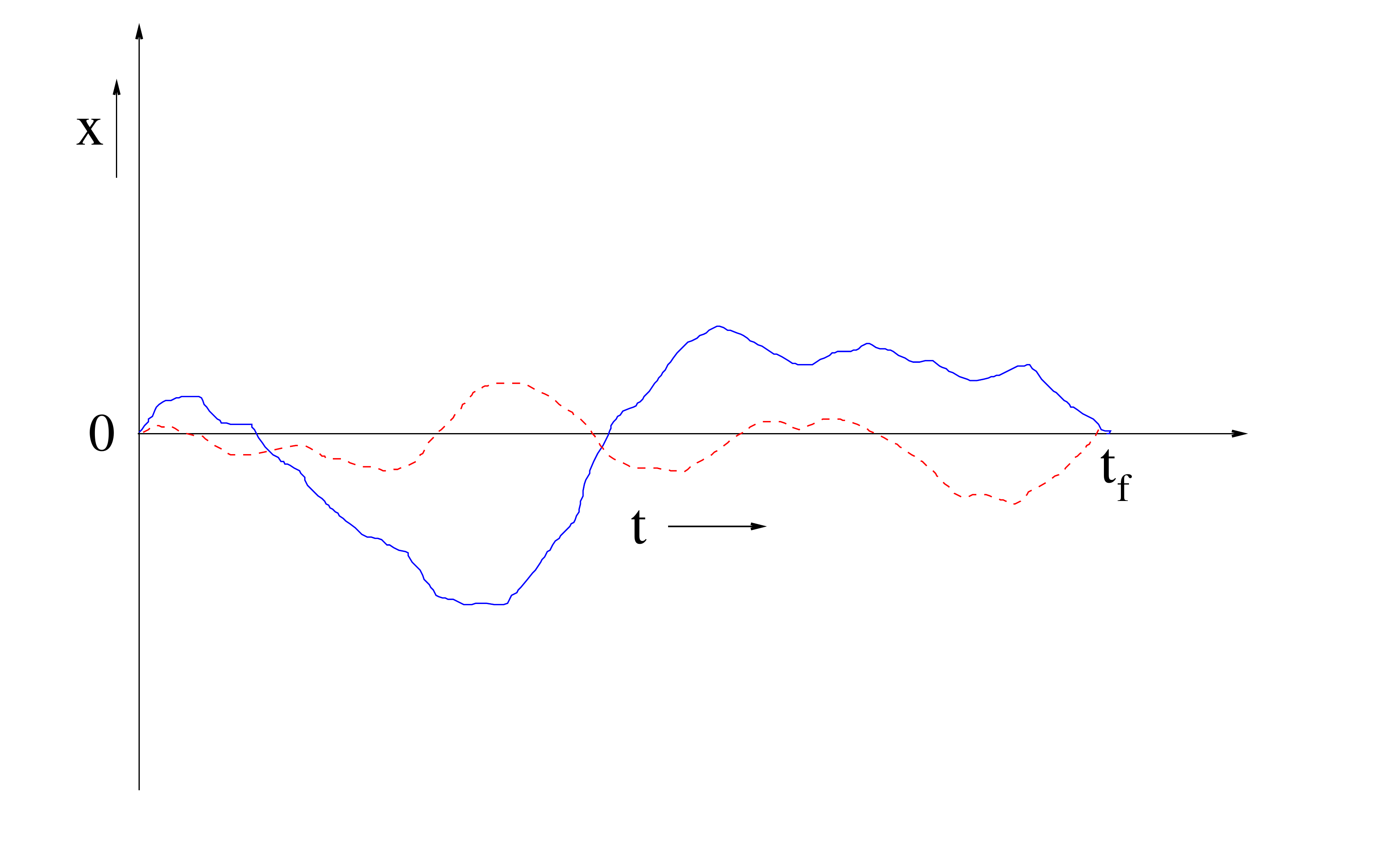}~~~
\includegraphics[width=.49\hsize]{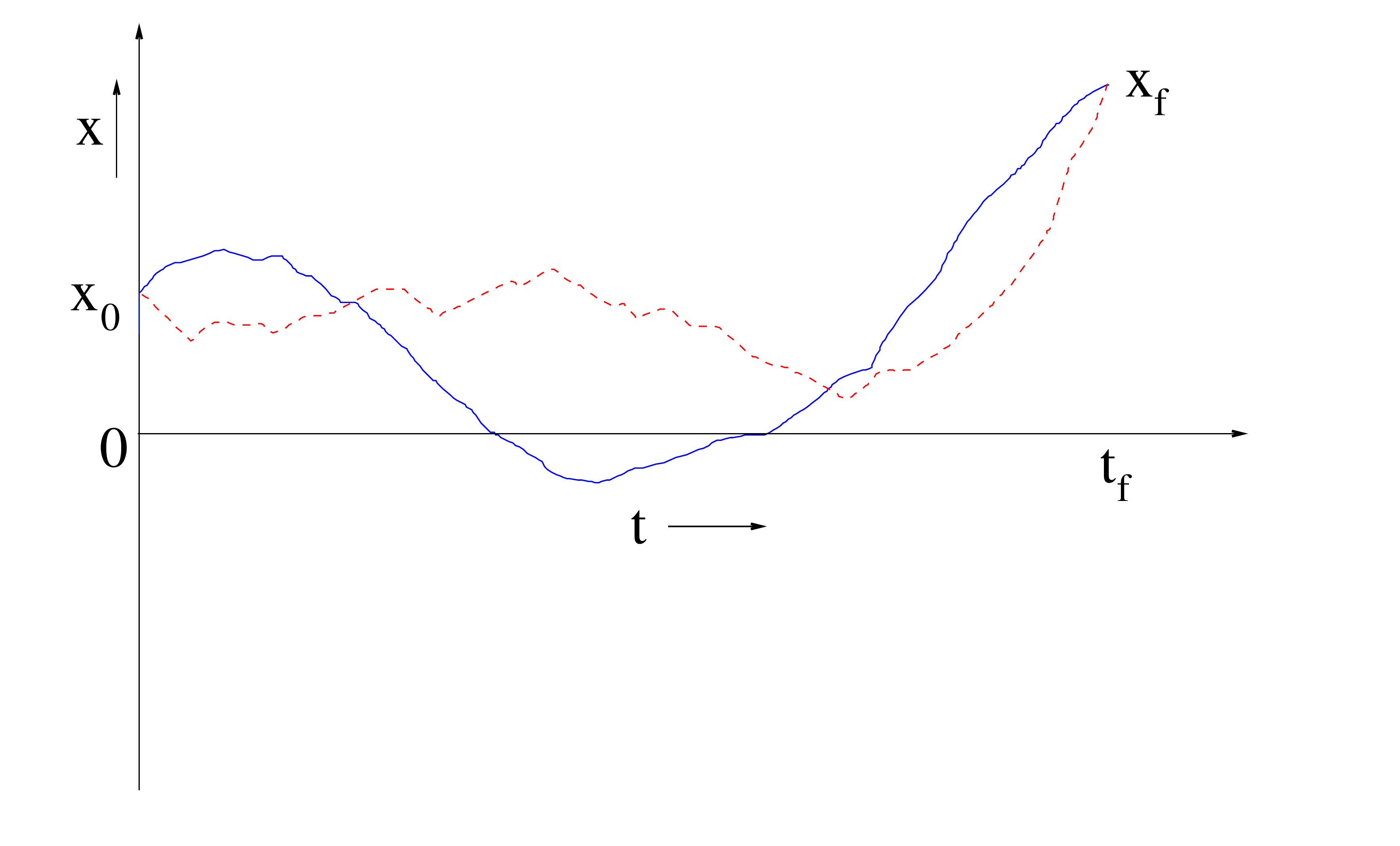}
\caption{(a) Two configurations of a Brownian bridge, starting at $x_B(t)=0$,  and returning to $x_B(t_f)=0$ at time $t_f$
(b) two configurations of a generalised Brownian bridge starting at $x_B(0)=x_0$ and
ending at the final position $x_B(t_f)=x_f$ at time $t_f$ ($x_f$ fixed, but not necessarily zero).}
\label{bridge.fig}
\end{figure}

\noindent {\em Brownian bridge:} A Brownian bridge $x_B(t)$ in $0\le t\le t_f$ is a Brownian motion, starting
at $x_B(0)=0$, that is constrained to come back to its starting point at a final time $t_f$, i.e., 
$x_B(t_f)=0$ (see
Fig. (\ref{bridge.fig}\, a)). A generalised Brownian bridge corresponds to those configurations
of Brownian motions that start at $x_B(0)=x_0$ and end at $x_B(t_f)=x_f$ (with $x_f$ fixed, not necessarily
the same as the starting position $x_0$) at time $t_f$ (see Fig. (\ref{bridge.fig}\,b)).

\vskip 0.3cm

\begin{figure}
\includegraphics[width=.49\hsize]{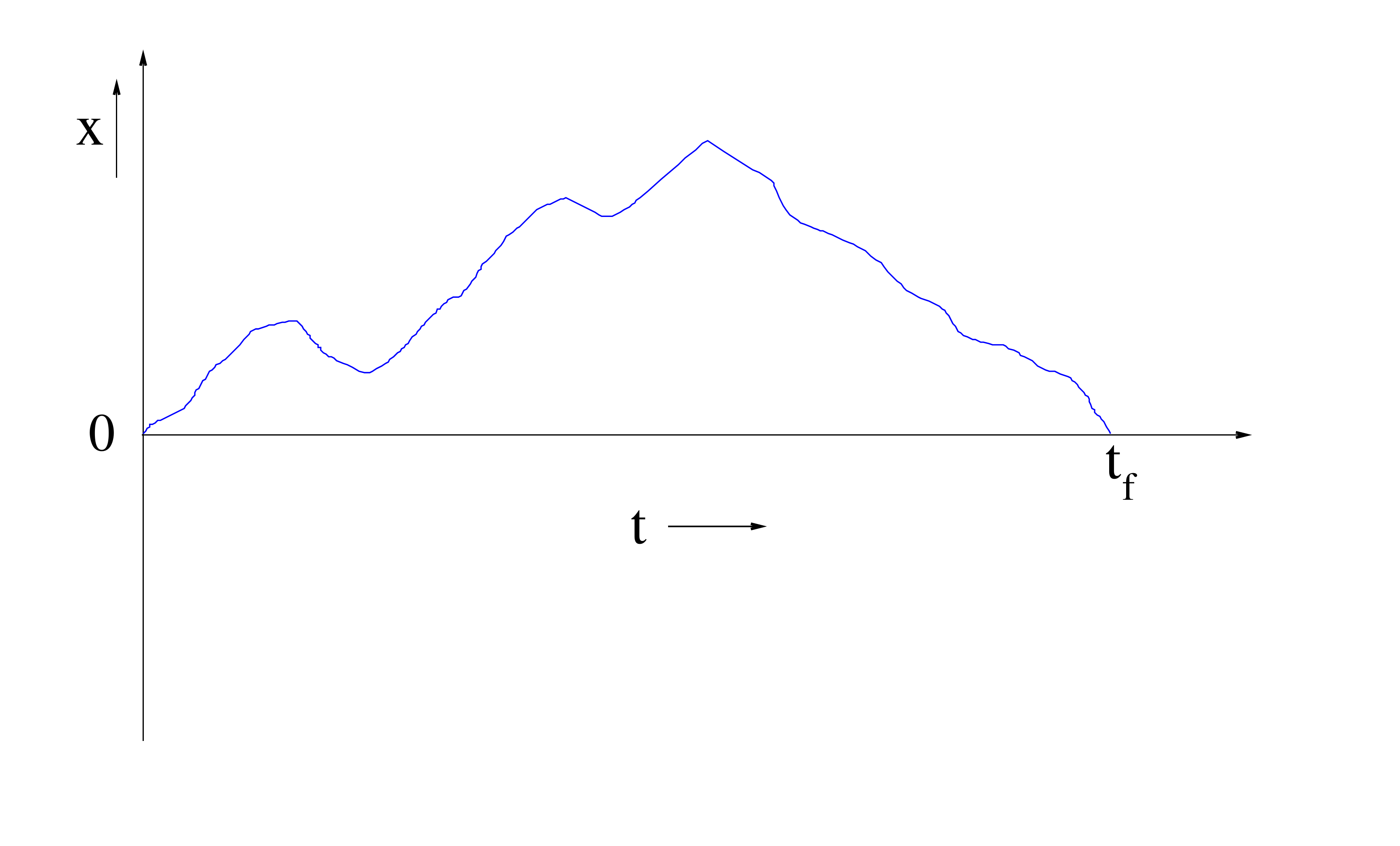}~~~
\includegraphics[width=.49\hsize]{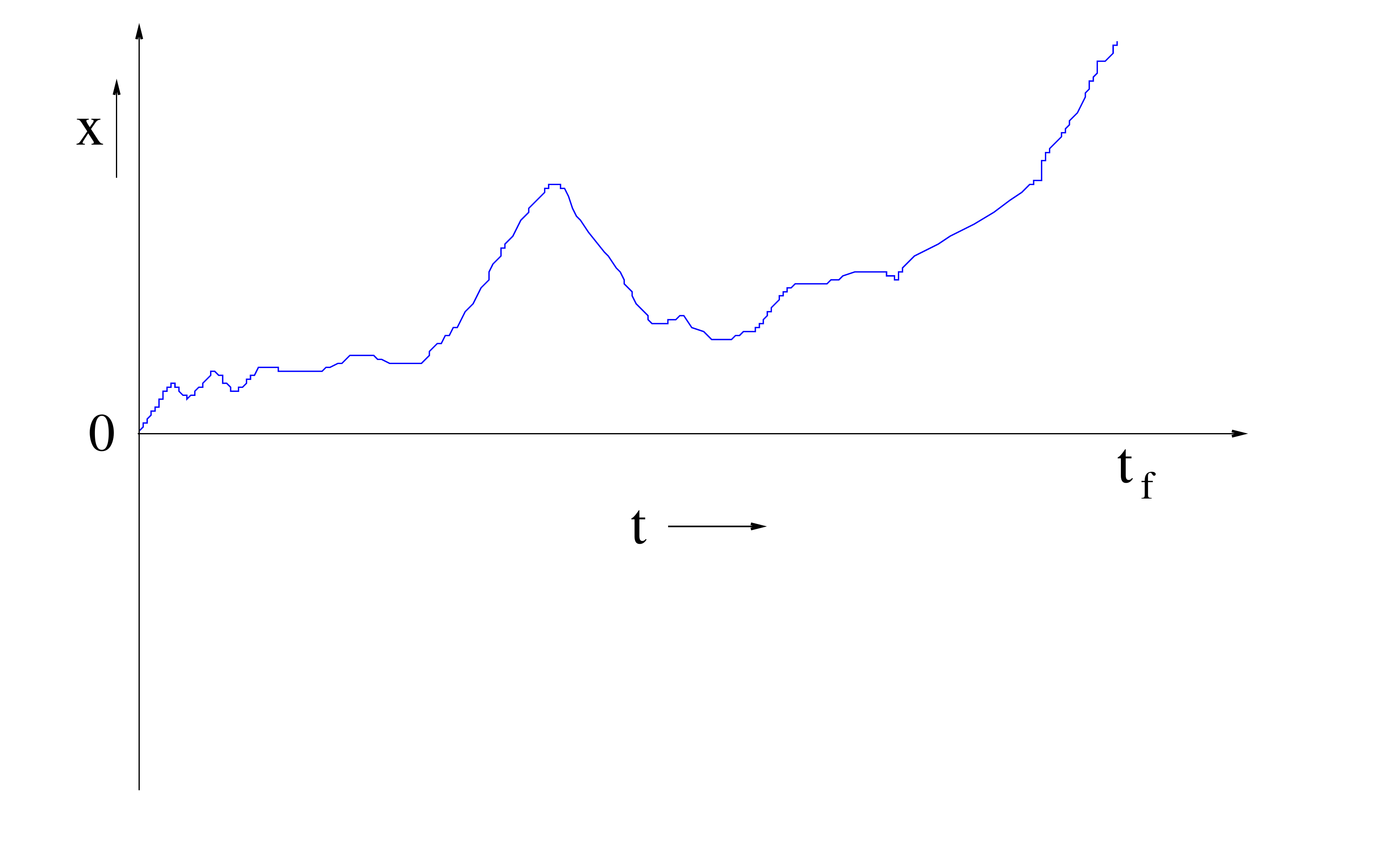}
\caption{(a) A typical configuration (schematic) of a Brownian excursion in the time interval $[0,t_f]$,
that starts at $x_E(0)=0$, ends at
$x_E(t_f)=0$, and stays positive in the whole interval $t\in [0,t_f]$ and  
(b) A typical configuration (schematic) of a Brownian meander in the time interval $[0,t_f]$,
that starts at $x_M(0)=0$, stays positive in the time period $[0,t_f]$, but its final position
$x_M(t_f)$ can be any positive number.} 
\label{excursion.fig}
\end{figure}

\noindent {\em Brownian excursion:} A Brownian excursion $x_{E}(t)$ in $0\le t\le t_f$ is a Brownian motion
that starts at $x_E(0)=0$, ends at $x_E(t_f)=0$ (as in a bridge), but additionally is constrained to stay positive
in between, i.e., for all $0<t<t_f$ (see Fig. (\ref{excursion.fig}\, a) for a typical excursion configuration).

\vskip 0.3cm

\noindent {\em Brownian meander:} A Brownian meander $x_M(t)$ is a Brownian motion in $[0,t_f]$, 
that starts at $x_M(0)=0$, is constrained to stay positive in the interval $t\in [0,t_f]$, but its
final position $x_M(t_f)$ is free, i.e., can be any positive number. For a typical meander configuration, see
Fig. (\ref{excursion.fig}\,b). 

A natural question is: how does one generate a constrained path such as a bridge, excursion or a meander
numerically with the correct statistical weight? Let us start with the example of a
simple Brownian bridge $x_B(t)$ over $[0,t_f]$ with the constraint $x_B(0)=x_B(t_f)=0$: how does one 
generate such a bridge?
A naive algorithm would be to generate all possible
configurations of a free
Brownian motion over $[0,t_f]$ via the discretized Langevin equation \eqref{brown_discrete} starting from
the origin and retain only those that do return to the origin at $t_f$. Evidently, such a naive algorithm 
is too wasteful and would not lead to correct statistics for a finite number of samples.
For the Brownian bridge, there is a however simple way out. One first generates a free Brownian path $B(t)$
via \eqref{brown_discrete} starting from $B(0)=0$ and from this path, one constructs a new path
$x_B(t)$ via the following simple construction
\begin{equation}
x_B(t)= B(t) - \frac{t}{t_f}\, B(t_f)\quad\, t\in [0,t_f]\, .
\label{bridge_alg.1}
\end{equation}  
By construction $x_B(t)$ in \eqref{bridge_alg.1} satisfies the bridge condition $x_B(0)=x_B(t_f)=0$.
In addition, by computing the propagator for the constructed process $x_B(t)$, it is easy to show that
indeed $x_B(t)$ has the correct statistical weight of a Brownian bridge.

\begin{figure}
\includegraphics[width=.49\hsize]{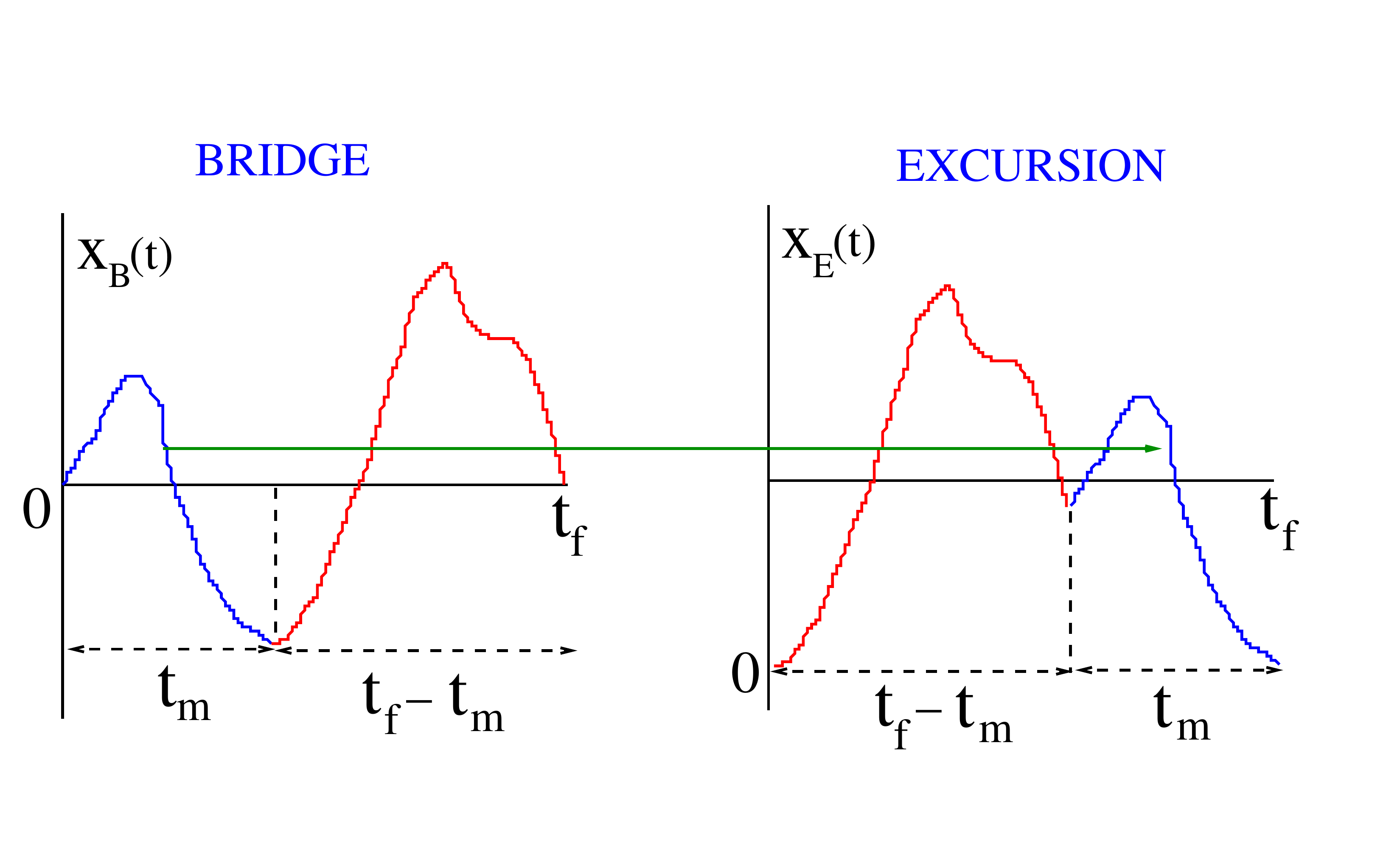}
\caption{Generation of an excursion from a bridge via Vervaat's construction.
One starts with a bridge configuration $x_B(t)$ with $x_B(0)=x_B(t_f)=0$ (left side).
The bridge achieves its minimum value $x_{\rm min}$ at time $t_m$ (shown by the
dashed vertical line). Divide the interval $[0,t_f]$ into two subintervals $[0,t_m]$ (blue)
and $[t_m,t_f]$ (red). Slide the left (blue) subinterval in the forward time direction
by an amount $t_f$ (shown by the (green) arrow) and join it to the right of the
subinterval $[t_m, t_f]$. Now shift the $x$ axis so that $x_{\rm min}$ denotes the new origin.
This newly constructed cut and paste bridge configuration over $t_m,t_m+t_f]$ is indeed
an excursion $x_E(t)$ over the time interval $[0,t_f]$.} 
\label{vervaat_fig}
\end{figure}

Similarly, one can generate an excursion $x_E(t)$, satisfying $x_E(0)=x_E(t_f)=0$ and
$x_E(t)$ remaining positive in between, from a bridge configuration $x_B(t)$ via the 
so called Vervaat construction~\cite{Vervaat} which proceeds as follows. First, we construct a bridge $x_B(t)$
from a free Brownian path using \eqref{bridge_alg.1}. Let $t_m$ denote the time at which
the bridge $x_B(t)$ achieves its minimum value say $x_{\rm min}$ (see Fig. (\ref{vervaat_fig})). We divide
the full time interval $[0,t_f]$ into two subintervals $[0,t_m]$ and $[t_m,t_f]$. Next 
we take the portion of the bridge path $x_B(t)$ in $[0,t_m]$ and slide it in the forward time 
direction by an interval $t_f$ [as shown in Fig. (\ref{vervaat_fig})] and paste it to
the other side of the rest of $x_B(t)$ in the second subinterval $[t_m,t_f]$. Due to
the periodicity of the bridge, the newly slided and pasted section of the bridge will join smoothly with
the part over $[t_m,t_f]$. Next, we shift the origin to $x_{\rm min}$, and the
resulting process over the new time-interval $[t_m, t_f+t_m]$ is an excursion over $[0,t_f]$.
In other words, from the bridge $x_B(t)$, we construct the excursion $x_E(t)$ via
the construction
\begin{equation}
x_E(t) = x_B(t_m+t)-x_{\rm min}\quad\, t\in [0,t_f]\, 
\label{excur_alg.1}
\end{equation}
where by periodicity $x_B(t+t_f)=x_B(t)$.
By construction in \eqref{excur_alg.1}, $x_E(0)=x_E(t_f)=0$ and $x_E(t)\ge 0$ for all $0<t<t_f$.
Moreover, Vervaat proved rigorously that the construction in \eqref{excur_alg.1} indeed
generates a Brownian  excursion with
the correct statistical weight~\cite{Vervaat}.

Another simpler way  to generate a Brownian excursion is to consider three independent
Brownian bridges $x_{B,1}(t)$, $x_{B,2}(t)$ and $x_{B,3}(t)$ and then construct a new process
\begin{equation}
x_E(t)= \sqrt{ [x_{B,1}(t)]^2 + [x_{B,2}(t)]^2 + [x_{B,3}(t)]^2}\, ,
\label{excur_alg.2}
\end{equation}
which just represents the radial part of a $3$-dimensional Brownian bridge ($3$-d Bessel bridge).
This new process can be rigorously shown to have the correct statistical weight of a one dimensional Brownian
excursion~\cite{Williams,Imhof}. Similarly, a Brownian meander $x_M(t)$ in one dimension, starting at
$x_M(0)=0$ and ending at $x_M(t_f)=x_f$ can be generated from a $3$-dimensional Bessel bridge 
by the following construction~\cite{Williams,Imhof}
\begin{equation}
x_M(t)= \sqrt{ [x_{B,1}(t)]^2 + [x_{B,2}(t)]^2 + [x_{B,3}(t) + \frac{t}{t_f}\, x_f ]^2 }\, .
\label{meander_alg.1}
\end{equation}
For a nice review on the connection between bridges, excursions
and meanders, see~\cite{Bertoin1,BPY} and for recent applications of such 
constrained paths, see~\cite{majumdar_review,MC,KM1,Svante,KM2,KMM,perret,moloney}.

These clever transformations connecting bridges, excursions and meanders to free Brownian motion
are, of course, very efficient to generate them numerically. However, these transformations
rely on the specific properties of a free Brownian motion and will, in general, not hold for
a more general process such as a particle diffusing in an external potential.
Moreover, one may want to generate paths which are not simply bridges, excursions or meanders, but
constrained in some other way. For example, in many applications in computer science, one
needs to generate, say a Brownian excursion $x_E(t)$ over $t\in [0,t_f]$ with a fixed area
under the excursion $A=\int_0^{t_f} x_E(t)\, 
dt$~\cite{Darling,Louchard,Flajolet,MC,majumdar_review,KM1,Svante,KM2,KMM}.
Hence, it would be much more desirable to build a recipe to construct an effective Langevin equation like
\eqref{brown_discrete} which (i) will automatically take into account the global constraints
and (ii) will hold for more general stochastic processes such as a particle diffusing in an
external potential. Once we have such a recipe for an effective Langevin equation, it
can subsequently be easily time-discretized leading to an efficient method to generate constrained paths.
The purpose of this paper is to provide precisely such a recipe and demonstrate it with several examples.
Some of the results to be discussed in this paper were probably known, at least partially, 
in the probability literature
but in a language perhaps not easily accessible to physicists. One of the purposes of this paper is
to unveil the recipe for an effective Langevin equation in a physicist friendly language.
For a recent review on some aspects of constrained stochastic processes, namely, constraints
on large deviations of a stochastic variable, see~\cite{CT}.

The rest of the paper is organized as follows: we first derive the effective Langevin equation which satisfies the boundary constraint. We then illustrate the method on four analytically soluble cases, namely the Brownian bridges, Brownian meanders, Brownian excursions and the  Ornstein-Uhlenbeck process. We then show how this equation can be solved exactly numerically in the case of any landscape potential $U(x)$.

\section{Derivation of effective Langevin equation}

From now on, we assume that the system is driven by a force $F(x,t)$ and is subject to stochastic dynamics
in the form of an overdamped Langevin equation.

For the sake of simplicity, we illustrate the method on a one-dimensional
system, the generalization to higher dimensions or larger number of
degrees of freedom being straightforward. We follow closely the presentation given in Ref. \cite{orland1}.

The over damped Langevin equation reads

\begin{equation}
\frac{dx}{dt}=\frac{1}{\gamma}F(x(t),t)+\eta(t)\label{eq:langevin}
\end{equation}
where $x(t)$
is the position of the particle at time $t$, driven by the force $F(x,t)$, $\gamma$
is the friction coefficient, related to the diffusion constant $D$
through the Einstein relation $D=k_{B}T/\gamma$, where $k_{B}$ is the Boltzmann
constant and $T$ the temperature of the thermostat. In addition,
$\eta(t)$ is a Gaussian white noise with moments given by

\begin{equation}
<\eta(t)>=0\label{eq:noise1}\end{equation}
 \begin{equation}
<\eta(t)\eta(t')>=\frac{2 k_{B}T}{\gamma}\delta(t-t')\label{eq:noise2}\end{equation}

The probability distribution $P(x,t)$ for the
particle to be at point $x$ at time $t$ satisfies a Fokker-Planck
equation \cite{11}

\begin{equation}
\frac{\partial P}{\partial t}=D\frac{\partial}{\partial x}\left(\frac{\partial P}{\partial x}-\beta F P\right)\label{eq:FP}
\end{equation}
where $\beta=1/k_{B}T$ is the inverse temperature. This equation is to be supplemented by the initial condition $P(x,0)=\delta(x-x_{0})$, where the particle is assumed to be at $x_0$ at time $t=0$. To emphasize this initial condition, we will often use the notation $P(x,t)= P(x,t|x_0,0)$.

We now study the probability over all paths starting at $x_{0}$ at
time $0$ and conditioned to end at a given point $x_{f}$ at time $t_{f}$, to find the particle
at point $x$ at time $t \in [0,t_f]$. This is the {\em generalized bridge}. This probability can be written as 
\[
\mathcal{P}(x,t)=\frac{1}{P(x_{f},t_{f}|x_{0},0)}Q(x,t)P(x,t)
\]
where we use the notation
\[
P(x,t)=P(x,t|x_{0},0)\]
\[
Q(x,t)=P(x_{f},t_{f}|x,t)\]

Indeed, the probability for a path starting from $(x_0,0)$ and ending at $(x_f,t_f)$ to go through $x$ at time $t$ is the product of the probability $P(x,t|x_{0},0)$ to start at $(x_0,0)$ and to end at $(x,t)$ by the probability 
$P(x_{f},t_{f}|x,t)$ to start at $(x,t)$ and to end at $(x_f,t_f)$.

The equation satisfied by $P$ is the Fokker-Planck equation mentioned above (\ref{eq:FP}), whereas
that for $Q$ is the so-called reverse or adjoint Fokker-Planck equation \cite{11} given by

\begin{equation}
\frac{\partial Q}{\partial t}=-D\frac{\partial^{2}Q}{\partial x^{2}}-D\beta F \frac{\partial Q}{\partial x}\label{eq:FTadj}\end{equation}

It can be easily checked that the conditional probability
$\mathcal{P}(x,t)$ satisfies a new Fokker-Planck equation

\[
\frac{\partial\mathcal{P}}{\partial t}=D\frac{\partial}{\partial x}\left(\frac{\partial\mathcal{P}}{\partial x}-\left(\beta F + 2\frac{\partial \ln Q}{\partial x}\right)\mathcal{P}\right)\]

Comparing this equation with the initial Fokker-Planck (\ref{eq:FP})
and Langevin (\ref{eq:langevin}) equations, one sees that it can
be obtained from a Langevin equation with an additional potential force

\begin{equation}
\frac{dx}{dt}=D\beta F+2D\frac{\partial\ln Q}{\partial x}+\eta(t)
\label{eq:bridge1}\end{equation}

This equation has been originally obtained by Doob \cite{17} in the probability literature 
 and is known as the Doob transform of the Langevin equation (\ref{eq:langevin}). It was also presented in the physics literature in the context of barrier crossing \cite{orland1}.
It provides a simple recipe to construct a \emph{generalized 
bridge}. It generates Brownian paths, starting at $(x_{i},0)$
conditioned to end at $(x_{f},t_{f})$, with unbiased statistics.
It is the additional term $2D\frac{\partial\ln Q}{\partial x}$ in the Langevin equation that guarantees that the
trajectories starting at $(x_{0},0)$ and ending at $(x_{f},t_{f})$ are statistically unbiased. Note that this additional term is a priori time dependent, and thus the effective force which conditions the paths is space and time dependent.

In the following, we will specialize to the case where the force $F$ is derived from a potential $U(x)$.
The bridge equation becomes
\begin{equation}
\frac{dx}{dt}=-\frac{D}{k_{B}T} \frac{\partial U}{\partial x}+2D\frac{\partial\ln Q}{\partial x}+\eta(t)
\label{eq:bridge2}
\end{equation}

In that case, the Fokker-Planck equation can be recast into an imaginary time Schr\"odinger equation \cite{11}, and the probability distribution function $P$ can be written as
\begin{equation}
\label{eq:matrix}
Q(x,t)=P(x_f,t_f|x,t) = e^{-\beta/2 (U(x_f)-U(x))} <x_f| e^{-H(t_f-t)}|x>
\end{equation}
where the "quantum Hamiltonian" $H$ is defined by
\begin{equation}
H= -D \frac{\partial^2}{\partial x^2} + D V(x)
\end{equation}
and the potential $V$ by
\begin{equation}
V= \left(\frac{\beta}{2} \frac{\partial U}{\partial x}\right)^2 -\frac{\beta}{2} \frac{\partial^2 U}{\partial x^2}
\end{equation}

We denote by $M$ the matrix element of the Euclidian Schr\"odinger evolution operator

\[
M(x,t)= <x_f| e^{-H(t_f-t)}|x>
\]

Using eq.(\ref{eq:matrix}) for $Q$, one can write equation (\ref{eq:bridge2})
as

\begin{equation}
\frac{dx}{dt}=2\frac{k_{B}T}{\gamma}\frac{\partial}{\partial x}\ln<x_{f}|e^{-(t_{f}-t)H}|x>+\eta(t)\label{eq:bridge3}\end{equation}

We see on the above form that when $t \to t_f$, the matrix element $M(x,t)$ converges to $\delta(x_f-x)$, and it is this singular attractive potential which drives all the paths to $x_f$ at time $t_f$.

Note that for large time $t_{f}$, $M(x,t)$ 
is dominated by the ground state of $H$, namely 
\begin{equation}
\label{GS}
<x_{f}|e^{-(t_{f}-t)H}|x> \sim e^{-E_0 (t_f -t)} \Psi_0(x_f) \Psi_0(x)
\end{equation}

It is well known that the ground state wave function of $H$ is 
$\Psi_0(x) = \frac{1}{\cal N} e^{-\beta U(x)/2}$ where ${\cal N}$ is the normalization constant. Indeed, it is easily checked that $H |\Psi_0\rangle =0$  which implies that $\Psi_0$ is an eigenstate of $H$ with ground state energy $E_0=0$. Inserting these relations in eq.(\ref{eq:bridge3}),
one recovers, as expected, the original unconditioned Langevin
equation for large time $t_f$.
Note that although relation (\ref{GS}) holds only for a normalizable ground state wave function $\Psi_0(x)$, it still yields the correct Langevin equation (\ref{eq:langevin}) for non-normalizable cases (such as the free Brownian motion $U(x)$=0, for which $E_0=0$ and $\Psi_0=1/\sqrt{V}$ where $V$ is the volume).

In order to build a representative sample of paths starting at $(x_{0},0)$
and ending at $(x_{f},t_{f}),$ one must simply solve equation (\ref{eq:bridge2})
for many different realizations of the random noise. 
Only the initial boundary condition is to be imposed,
as the singular term in the equation imposes the correct final boundary
condition. An important point to note is that all the trajectories generated by this procedure are statistically independent.

At this stage, we see that all the difficulty lies in calculating either the Green's function $Q(x,t)$ or the matrix element  $M(x,t)$. In general, this is not possible analytically except for a few special cases. In ref.\cite{orland1}, a short time approximation to compute $M(x,t)$ was presented, but in the following, we specialize on a few exactly analytically solvable cases.

\section{The free Brownian Bridge}

The simplest case is the case of a free particle in unrestricted space. In that case, the Green's function $Q(x,t)$ can be easily computed and yields
\[
Q(x,t)=\frac{1}{\sqrt{4 \pi D (t_f-t)}} e^{-\frac{(x_f-x)^2}{4 D (t_f-t)}}
\]
The corresponding conditioned Langevin equation becomes
\begin{equation}
\label{free}
\frac{dx}{dt}= \frac{x_f-x}{t_f-t} + \eta(t)
\end{equation}

\begin{figure}
\includegraphics[width=.7\hsize]{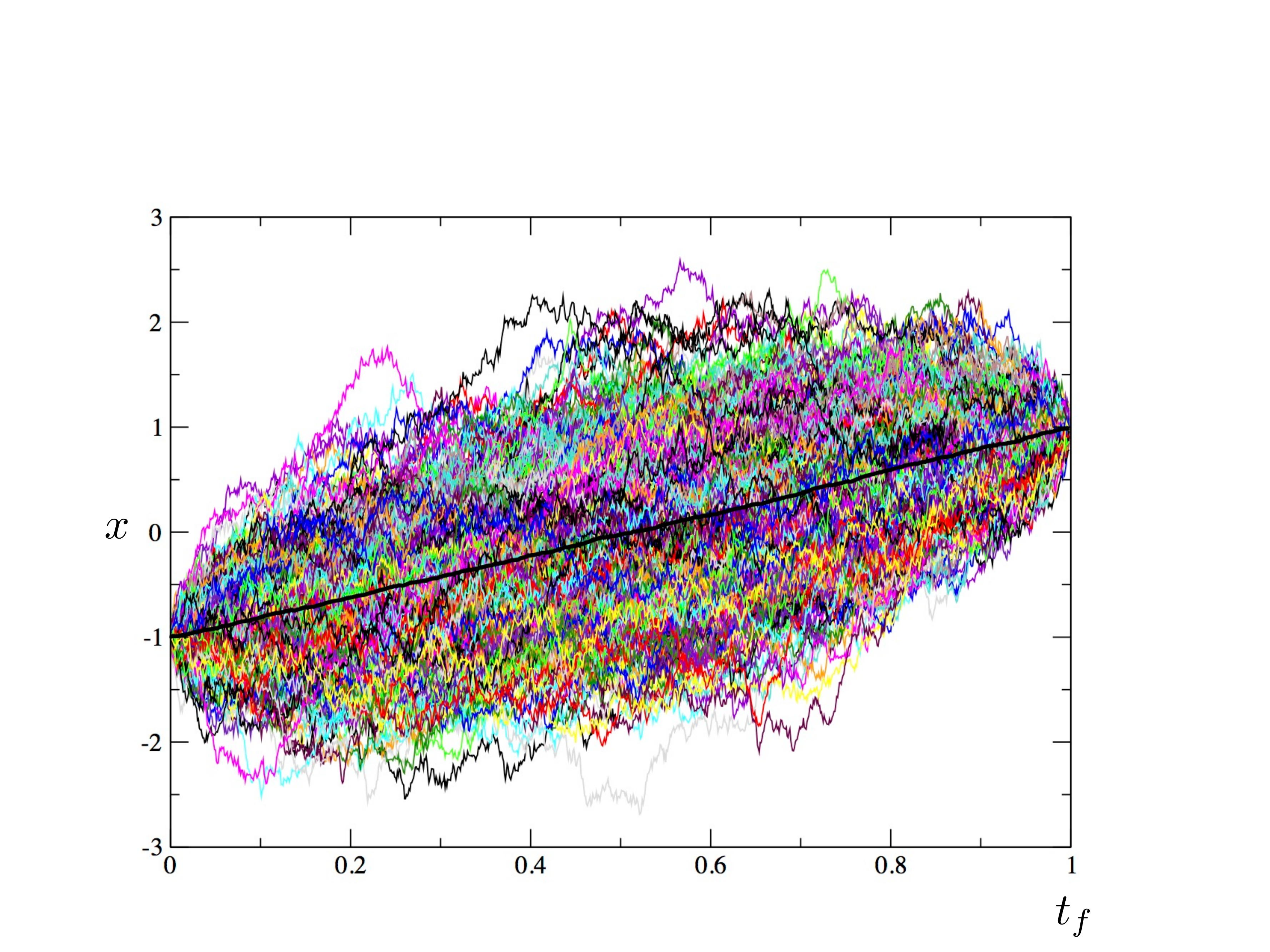}
\caption{A sample of 500 Brownian bridges} 
\label{fig1}
\end{figure}

This equation \eqref{free} can be found in \cite{rogers_williams,CT}. It is completely universal as the drift term does not 
depend on the diffusion 
constant $D$. We show in Fig.(\ref{fig1}) a set of 500 trajectories starting at $x_0=-1$ at time $0$ and ending at $x_f=+1$ at time $t_f=1$, obtained for different noise histories. The time step used in the discretization is $dt=0.001$. All these trajectories are statistically independent. The thick black curve is the mean trajectory. As can be seen from eq.(\ref{free}), it is a straight line.

\section{The Brownian meander and excursion}
The Brownian meander is a Brownian path of a free particle in a half-space.
In other words, the Brownian particle cannot penetrate in one of the half-planes. 
In 1d, if we define the half-space as $x>0$, the particle cannot go into the $x<0$ subspace.
Here the meander $x_M(t)$ starts, say at the origin $x_M(t=0)=x_0>0$, ends at $x_M(t_f)=x_f>0$, while
staying positive in $t\in [0,t_f]$. For the special case where both the initial and the final position
tend to zero, the meander is called an excursion $x_E(t)$. These constrained Brownian paths have been shown to be relevant to the anomalous diffusion of cold atoms in optical lattices \cite{barkai}.

In the case of a general meander, the Green's function $Q(x,t)$ is that of a particle restricted to a half space. This Green's function is well known and can be calculated by using the method of images. In the case when the end point is fixed at $x_f$ at time $t_f$ we obtain
\[
Q(x,t)=\frac{1}{\sqrt{4 \pi D (t_f-t)}} \left(e^{-\frac{1}{4D} \frac{(x_f-x)^2}{t_f-t}} - e^{-\frac{1}{4D} \frac{(x_f+x)^2}{t_f-t}} \right)
\]

If the end point $x_f$ at $t_f$ can be anywhere in the 1/2 plane $x_f>0$, the Green's function above has to be further integrated for $x_f \in [0,+\infty]$ to obtain
\begin{eqnarray}
Q_M(x,t)&=& \int_0^{\infty} dx_f \ Q(x,t) \nonumber \\
&=& {\rm erf}\left(\frac{x}{\sqrt{4 D (t_f-t)}}\right)
\end{eqnarray}
where ${\rm erf}(x)= \frac{2}{\sqrt{\pi}}\, \int_0^x e^{-u^2}\, du$.
The corresponding Langevin equation for the meander then reads
 \begin{equation}
 \label{meander}
 \frac{dx}{dt}= \frac{2}{\sqrt{4 \pi D (t_f-t)}} \ \frac{\exp \left( -\frac{x^2}{4 D (t_f-t)} 
\right)}{{\rm erf}(\frac{x}{\sqrt{4 D (t_f-t)}})}
 +\eta(t)
\end{equation}
 
The case of a Brownian excursion, where the extremity $x_f$ is fixed, is generated by the Langevin equation
 \begin{equation}
 \label{exc}
 \frac{dx}{dt}=\frac{
\left(\frac{x_f-x}{t_f-t} \right)
e^{-\frac{1}{4D} \frac{(x_f-x)^2}{t_f-t}} 
+ \left( \frac{x_f+x}{t_f-t}\right)
e^{-\frac{1}{4D} \frac{(x_f+x)^2}{t_f-t}} }{e^{-\frac{1}{4D} \frac{(x_f-x)^2}{t_f-t}} - e^{-\frac{1}{4D} \frac{(x_f+x)^2}{t_f-t}} }+\eta(t)
\end{equation}

Taking $x_f\to 0$, one obtains the effective Langevin equation for an excursion
 \begin{equation}
 \label{excursion}
 \frac{dx}{dt}= \frac{2D}{x} \left( 1 - \frac{x^2}{2D (t_f-t)} \right)
 +\eta(t)
\end{equation}

This equation can be found in ref.\cite{rogers_williams}.

\begin{figure}
\includegraphics[width=.7\hsize]{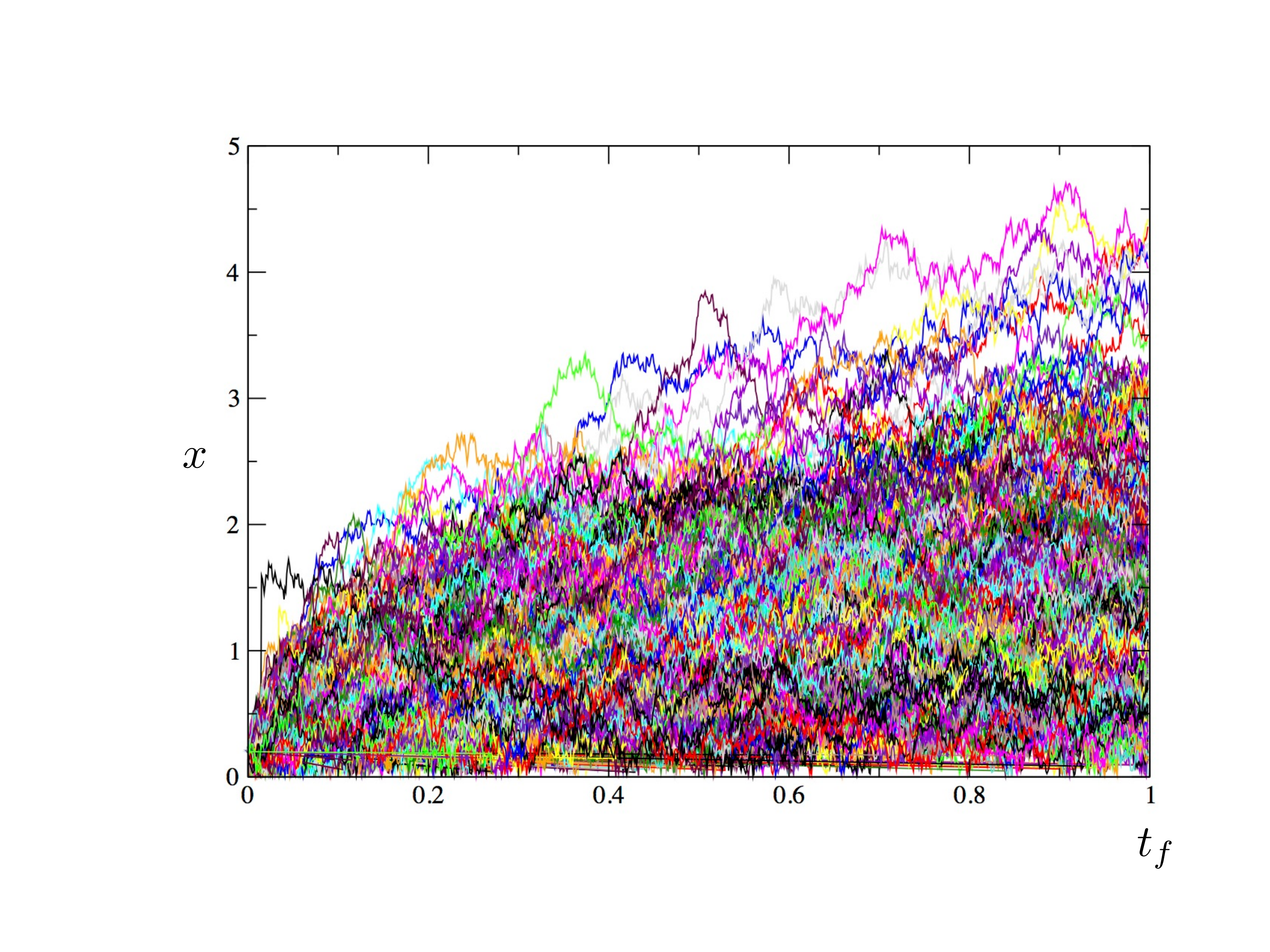}
\caption{A sample of 500 meanders} 
\label{fig2}
\end{figure}

\begin{figure}
\includegraphics[width=.7\hsize]{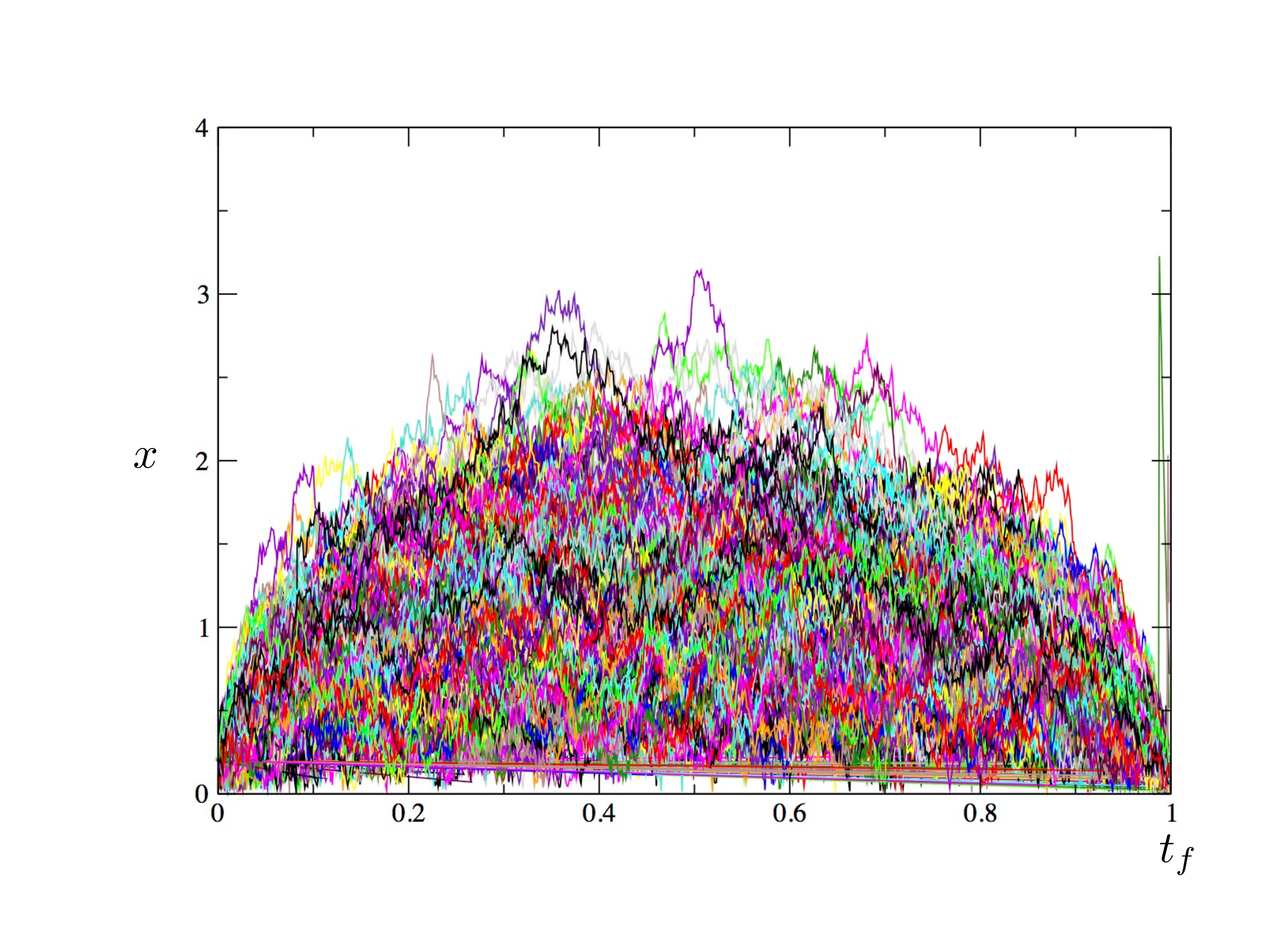}
\caption{A sample of 500 excursions} 
\label{fig3}
\end{figure}

Again, these two equations (\ref{meander}) and (\ref{excursion}) can be solved easily by discretizing time. We show in Fig.(\ref{fig2}) a set of 500 statistically independent meanders starting at $x_0=5$ at time $0$ and ending anywhere in the upper half plane at time $t_f=1$. 
Similarly, we show in Fig.(\ref{fig3}) a set of 500 statistically independent excursions starting
at $x_0=0.01$ at time $0$ and ending at $x_f=0$ at time $t_f=1$. In all cases, the independent trajectories are 
obtained for different noise histories and
time step $dt=0.001$. Note that generating an excursion by Eq. (\ref{excursion}) is
evidently much simpler than generating it by the Vervaat construction~\cite{Vervaat} discussed 
in
the introduction.

\section{The harmonic oscillator: the Ornstein-Uhlenbeck process}
Another solvable case is  that of the harmonic oscillator
$U(x)=K x^2/2$. The unconditioned Langevin equation is
\begin{equation}
\frac{dx}{dt}= -D\beta K x + \eta(t)
\end{equation}
This is a so-called Ornstein-Uhlenbeck process \cite{11}. The matrix element $M(x,t)$ associated to this potential can be computed easily and one obtains
\begin{equation}
M(x,t) =\frac{1}{\cal N} \exp{\Bigg( 
- \frac{K \beta}{ 4 \sinh (DK\beta \tau)} \Big( (x_f^2+x^2)\cosh (D K \beta \tau) -2 x x_f \Big) \Bigg)}
\end{equation}
where $\tau=t_f -t $ and ${\cal N}$ is a normalization constant.

The associated conditioned Langevin equation is thus
\begin{equation}
\label{OU}
\frac{dx}{dt}= DK\beta\ \frac{x_f-x \cosh DK\beta (t_f-t)}{\sinh DK\beta (t_f-t)} + \eta(t)
\end{equation}
which was obtained in \cite{CT}.

\begin{figure}
\includegraphics[width=.7\hsize]{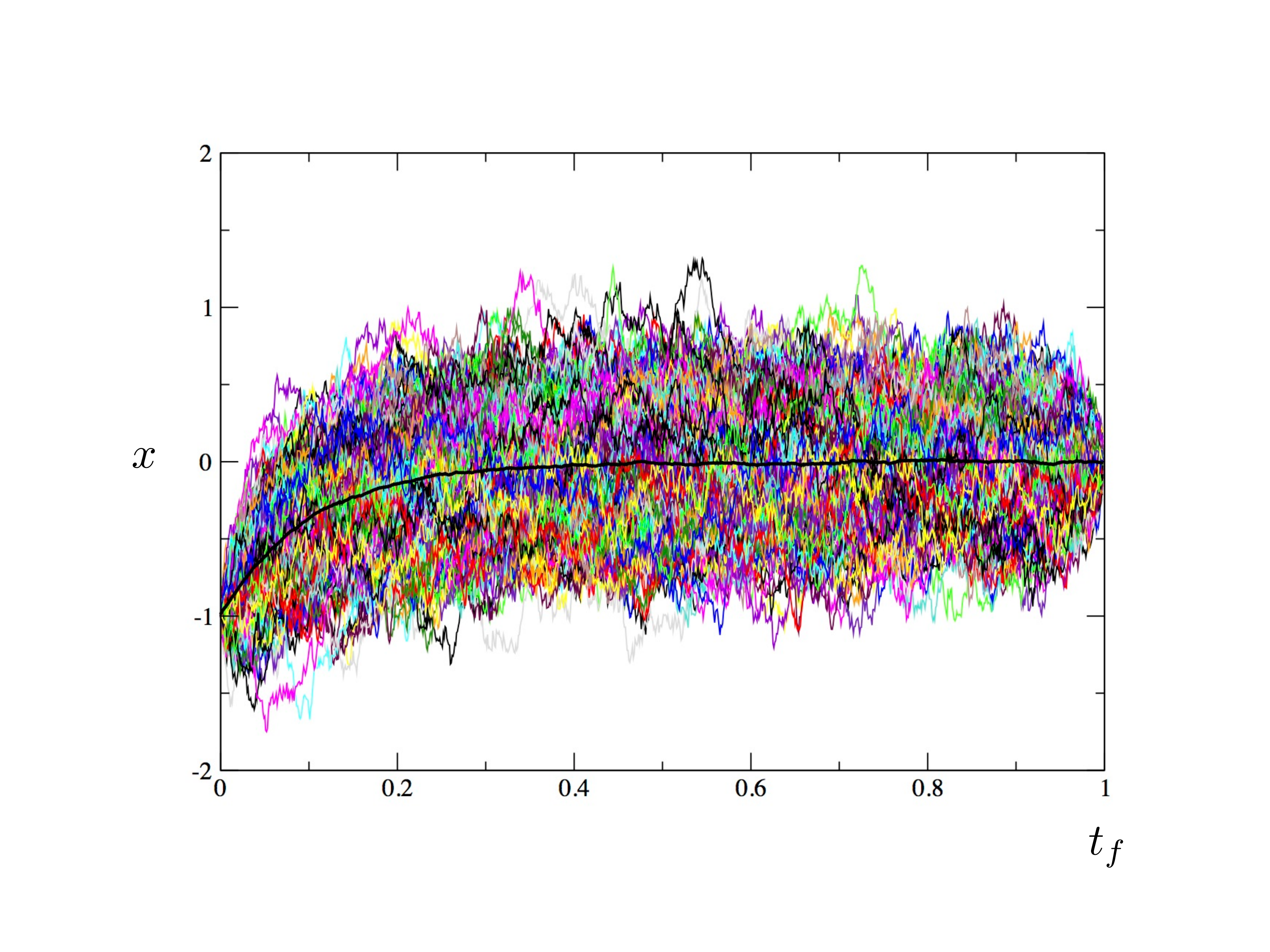}
\caption{A sample of 500 Ornstein-Uhlenbeck trajectories starting at -1 and ending at 0} 
\label{fig4}
\end{figure}

Note that this equation is invariant when changing $K$ into $-K$ and thus we have this surprising result that paths going up a barrier are statistically identical to those going down a barrier. To illustrate this apparent paradox, we show in Fig.(\ref{fig4}) a set of 500 trajectories starting at $x_0=-1$ at time $0$ and ending at $x_f=0$ at time $t_f=1$, obtained for different noise histories with $dt=0.001$. The simulation was performed at temperature $T=0.1$ with $K=1$.

\section{The general case}
There are very few cases which are analytically solvable and we thus now discuss the generic non-solvable case.
In order to compute the central quantity $M(x,t)$, we perform a spectral expansion \cite{12}
\begin{eqnarray}
M(x,t) &=& \langle x_f|e^{-(t_f-t)H}|x \rangle \nonumber \\
&=& \sum_ \alpha e^{-E_\alpha(t_f-t)}\Psi_{\alpha}(x_f) \Psi_{\alpha}(x)
\end{eqnarray}
where $\Psi_{\alpha}$ is the eigenstate of $H$ with eigenvalue $E_{\alpha}$
\[
H \Psi_\alpha (x) = E_\alpha  \Psi_(x)
\]

Note that although this spectral expansion is in principle valid only for normalizable eigenstates, it can easily be extended to the non-normalizable case by taking the limit of infinite volume.

The bridge equation (\ref{eq:bridge3}) becomes
\begin{equation}
\frac{dx}{dt}  = 2 D \frac{\sum_ \alpha e^{-E_\alpha(t_f-t)}\Psi_{\alpha}(x_f) \frac{\partial \Psi_{\alpha}(x)}{\partial x}}{\sum_ \alpha e^{-E_\alpha(t_f-t)}\Psi_{\alpha}(x_f) \Psi_{\alpha}(x)}+\eta(t)
\end{equation}
where we have used $D=k_B T/\gamma$.

In order to be able to solve this equation numerically (by discretization for instance), one has to compute the eigenstates $\Psi_{\alpha}$ and eigenvalues $E_{\alpha}$ of the Hamiltonian $H$. In the case of low-dimensional system, this can be done very easily by diagonalizing the discretized Hamiltonian $H$ which turns out to be a tridiagonal operator.

\section{The quartic double-well}
\label{double}

We illustrate the above method on the example of barrier crossing in
1d (quartic potential).

\[
U(x)=\frac{1}{4}(x^{2}-1)^{2}
\]

This potential has two minima at $x=\pm1$, separated by a barrier
of height 1/4. Note that $V(x)= (\beta U'/2)^2- \beta U''/2$
is much steeper than $U(x)$ and thus more confining around its minima.
At low temperature, the potential
$V(x)$ has two minima at points close to $\pm1$ and one minimum
at $x=0$.

The ground state of the Hamiltonian is $\Psi_0(x)\sim \exp(-\beta U(x)/2$.
The Hamiltonian is diagonalized by discretizing space and writing it in the form of a tridiagonal matrix. The diagonalization is performed using programs specific to tridiagonal matrices. All the examples were performed
at low temperature $T=0.05$, where the barrier height is equal to
5 in units of $k_{B}T$ and the Kramers relaxation time, given by
the inverse of the smallest non-zero eigenvalue of $H$, is equal
to $\tau_{K}=362.934$ .

\begin{figure}
\includegraphics[width=.7\hsize]{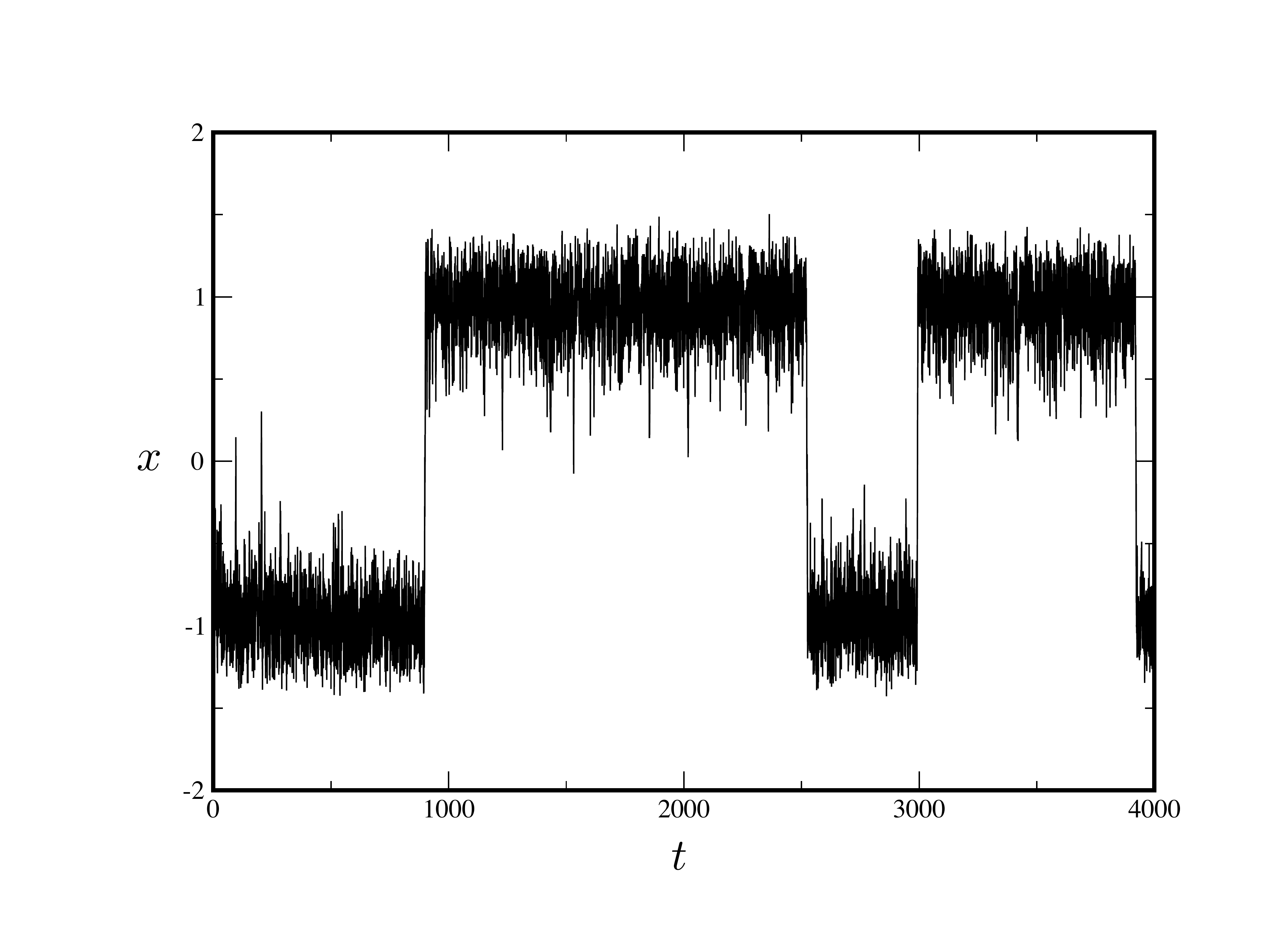}
\caption{A sample long unconstrained Langevin trajectory in the double-well.} 
\label{fig88}
\end{figure}

On fig.\ref{fig88}, we present a long trajectory
($t_{f}=1000)$ obtained by solving the unconditioned Langevin eq.(\ref{eq:langevin})
for a particle starting at $x_{0}=-1$ at time 0 . 

The general pattern is that of the particle
staying in one of the wells for a long time, then crossing
very rapidly into the other well and back and forth. If one is interested more precisely in the transition region, our method allows to perform the simulation of the rare crossing events between the two wells and generate a large sample of statistically independent paths in this transition region.

\begin{figure}
\includegraphics[width=.7\hsize]{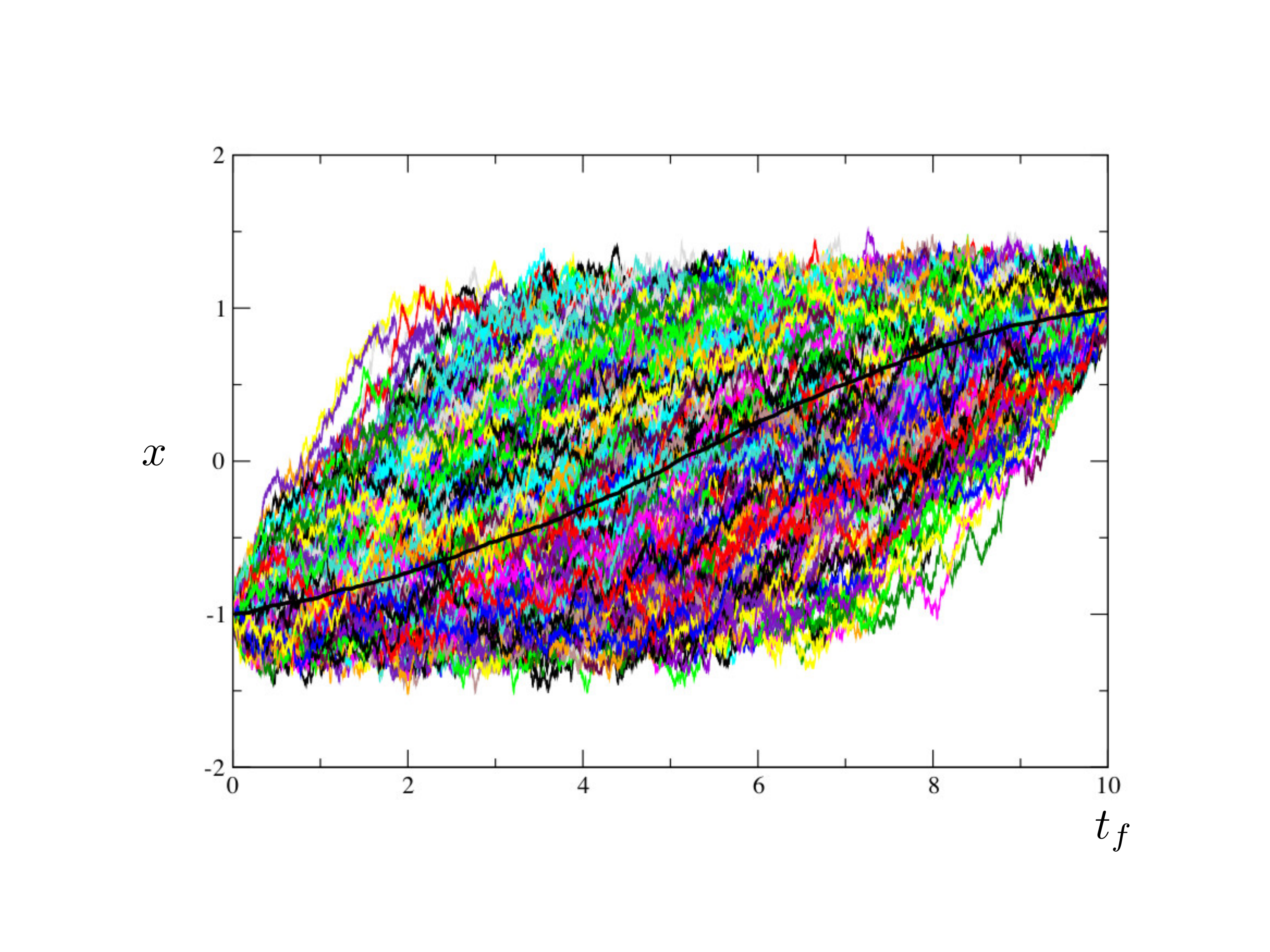}
\caption{A sample of 500 trajectories joining the two minima of the double-well.} 
\label{fig8}
\end{figure}

\begin{figure}
\includegraphics[width=.7\hsize]{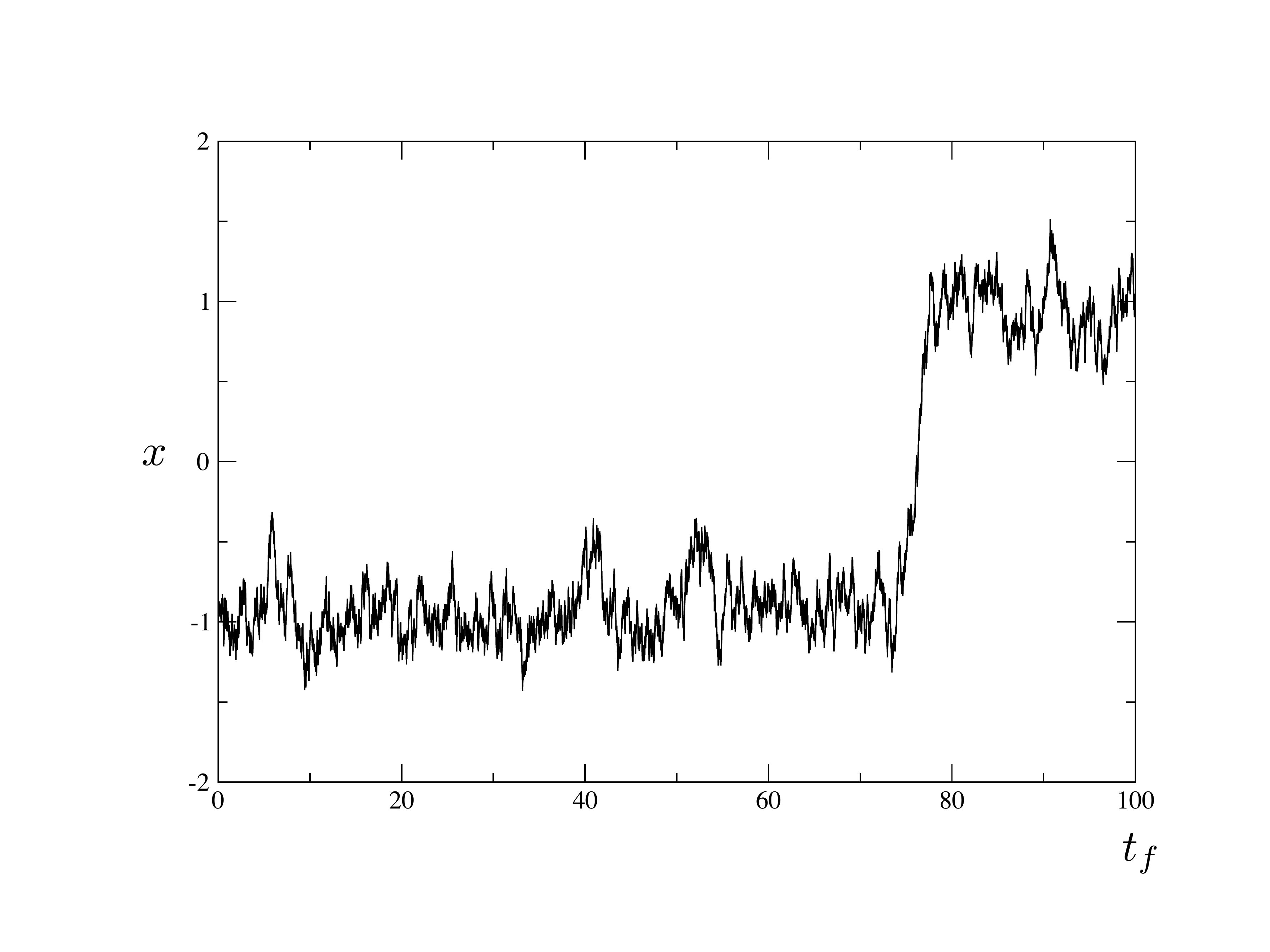}
\caption{A sample long trajectory in the double-well joining the two minima.} 
\label{fig9}
\end{figure}

In fig.\ref{fig8}, we show
a set of 500 trajectories starting at $x_0=-1$ at time $0$ and ending at $x_f=+1$ at time $t_f=10$, obtained for different noise histories with $dt=0.001$.
In fig.\ref{fig9}, we show a longer trajectory starting at $x_0=-1$ at time $0$ and ending at $x_f=+1$ at time $t_f=100$ with $dt=0.001$. 
%


\section{Conclusion}

We have presented in this paper a novel method in Physics to generate paths satisfying
the Langevin overdamped dynamics, starting from an initial point
and conditioned to end at a given final point at a given time. One of the great advantages of this method is that trajectories are generated by a modified Langevin equation and are all statistically independent. In this paper, we emphasized applications to analytically solvable models, such as bridges, meanders, excursions, Ornstein-Uehlenbeck processes, etc. In a forthcoming work, we will show how to generate paths with additional constraints, such as fixed area under the curve, fixed dissipation, etc.

\begin{acknowledgements}
We thank the hospitality of the Galileo Galilei Institute (Florence) during the workshop ``Advances in Nonequilibrium Statistical Mechanics: large deviations and long-range correlations, extreme value statistics, anomalous transport and long-range interactions" (May-June, 2014), where this work was initiated.
\end{acknowledgements}

\end{document}